\documentstyle[12pt]{article}

\newcommand{\nwc}{\newcommand}
\nwc{\hyp} {\hyphenation}
\nwc{\be}  {\begin{equation}}
\nwc{\ee}  {\end{equation}}
\nwc{\ba}  {\begin{array}}
\nwc{\ea}  {\end{array}}
\nwc{\bdm} {\begin{displaymath}}
\nwc{\edm} {\end{displaymath}}
\nwc{\bea} {\be\ba{lcl}}
\nwc{\eea} {\ea\ee}
\nwc{\bda} {\bdm\ba{lcl}}
\nwc{\eda} {\ea\edm}
\nwc{\bc}  {\begin{center}}
\nwc{\ec}  {\end{center}}
\nwc{\ds}  {\displaystyle}
\nwc{\bmat}{\left(\ba}
\nwc{\emat}{\ea\right)}
\nwc{\non} {\nonumber}
\nwc{\bib} {\bibitem}
\nwc{\lra} {\longrightarrow}
\nwc{\ra}  {\rightarrow}
\nwc{\Ra}  {\Rightarrow}
\nwc{\lmt} {\longmapsto}
\nwc{\prl} {\partial}
\nwc{\iy}  {\infty}
\nwc{\ol}  {\overline}
\nwc{\hm}  {\hspace{3mm}}
\nwc{\lf}  {\left}
\nwc{\ri}  {\right}
\nwc{\lm}  {\limits}
\nwc{\lb}  {\lbrack}
\nwc{\rb}  {\rbrack}
\nwc{\ov}  {\over}
\nwc{\pr}  {\prime}
\nwc{\nnn} {\nonumber \vspace{.2cm} \\ }
\nwc{\Sc}  {{\cal S}}
\nwc{\Lc}  {{\cal L}}
\nwc{\Rc}  {{\cal R}}
\nwc{\Dc}  {{\cal D}}
\nwc{\Oc}  {{\cal O}}
\nwc{\Cc}  {{\cal C}}
\nwc{\Pc}  {{\cal P}}
\nwc{\Mc}  {{\cal M}}
\nwc{\Ec}  {{\cal E}}
\nwc{\Fc}  {{\cal F}}
\nwc{\Hc}  {{\cal H}}
\nwc{\Kc}  {{\cal K}}
\nwc{\Xc}  {{\cal X}}
\nwc{\Gc}  {{\cal G}}
\nwc{\Zc}  {{\cal Z}}
\nwc{\Nc}  {{\cal N}}
\nwc{\fca} {{\cal f}}
\nwc{\xc}  {{\cal x}}
\nwc{\Ac}  {{\cal A}}
\nwc{\Bc}  {{\cal B}}
\nwc{\Uc}  {{\cal U}}
\nwc{\Vc}  {{\cal V}}
\nwc{\Th} {\Theta}
\nwc{\th} {\theta}
\nwc{\vth} {\vartheta}
\nwc{\eps}{\epsilon}
\nwc{\si} {\sigma}
\nwc{\Gm} {\Gamma}
\nwc{\gm} {\gamma}
\nwc{\bt} {\beta}
\nwc{\La} {\Lambda}
\nwc{\la} {\lambda}
\nwc{\om} {\omega}
\nwc{\Om} {\Omega}
\nwc{\dt} {\delta}
\nwc{\Si} {\Sigma}
\nwc{\Dt} {\Delta}
\nwc{\al} {\alpha}
\nwc{\vph}{\varphi}
\def\tr{\mathop{\rm tr}}

\def\VEV#1{\left\langle #1\right\rangle}

\def\pr#1{#1^\prime}

\nwc{\Id}  {{\bf 1}}
\nwc{\diag} {{\rm diag}}
\nwc{\inv}  {{\rm inv}}
\nwc{\mod}  {{\rm mod}}
\nwc{\hal} {\frac{1}{2}}
\nwc{\tpi}  {2\pi i}

\def\np#1{Nucl. Phys. {\bf B#1}}
\def\pl#1{Phys. Lett. {\bf B#1}}

\def\prd#1{Phys. Rev. {\bf D#1 }}
\def\mpl#1{Mod. Phys. Lett. {\bf A#1}}
\def\prle#1{Phys. Rev. Lett. {\bf #1}}

\newsavebox{\nnin} \sbox{\nnin}{$\hspace{1mm}\in\kern -.8em /
                   \hspace{1mm}$}

\newcommand{\sub}{\subset}
\newsavebox{\nnsub} \sbox{\nnsub}{$\hspace{1mm}\sub\kern -.9em /
            \hspace{1mm}$}

\def\KK{{\rm I\kern -.2em  K}}
\def\NN{{\rm I\kern -.16em N}}
\def\RR{{\rm I\kern -.2em  R}}
\def\ZZ{Z \kern -.43em Z}
\def\QQ{{\rm \kern .25em
             \vrule height1.4ex depth-.12ex width.06em\kern-.31em Q}}
\def\CC{{\rm \kern .25em
             \vrule height1.4ex depth-.12ex width.06em\kern-.31em C}}
\def\ZZZ{Z\kern -0.31em Z}

\catcode`\@=11

\def\@normalsize{\@setsize\normalsize{15pt}\xiipt\@xiipt
\abovedisplayskip 14pt plus3pt minus3pt%
\belowdisplayskip \abovedisplayskip
\abovedisplayshortskip  \z@ plus3pt%
\belowdisplayshortskip  7pt plus3.5pt minus0pt}

\def\small{\@setsize\small{13.6pt}\xipt\@xipt
\abovedisplayskip 16pt plus3pt minus3pt%
\belowdisplayskip \abovedisplayskip
\abovedisplayshortskip  \z@ plus3pt%
\belowdisplayshortskip  7pt plus3.5pt minus0pt
\def\@listi{\parsep 4.5pt plus 2pt minus 1pt
            \itemsep \parsep
            \topsep 9pt plus 3pt minus 3pt}}

\def\underline#1{\relax\ifmmode\@@underline#1\else
	$\@@underline{\hbox{#1}}$\relax\fi}
\@twosidetrue

\catcode`@=12

\catcode`\@=11

\def\thesection{\Roman{section}.}

\def\FERMIPUB{}
\def\FERMILABPub#1{\def\FERMIPUB{#1}}
\def\ps@headings{\def\@oddfoot{}\def\@evenfoot{}
\def\@oddhead{\hbox{}\hfill
	\makebox[.5\textwidth]{\raggedright\ignorespaces --\thepage{}--
	\hfill {\rm FERMILAB--Pub--\FERMIPUB}}}
\def\@oddhead{\hbox{}\hfill --\thepage{}-- \hfill}
\def\@evenhead{\@oddhead}
\def\subsectionmark##1{\markboth{##1}{}}
}

\ps@headings

\catcode`\@=12

\relax

\newcounter{appendix}

\def\appendix{\par
 \addtocounter{appendix}{1}
 \def\thesection{Appendix \Alph{appendix}:}
 \def\ksection{\Alph{appendix}}}

\begin{document}
\par \vskip .05in
\FERMILABPub{93/035--T}
\begin{titlepage}
\begin{flushright}
SSCL--Preprint--490\\
FERMI--PUB--93/035--T\\
July 29, 1993
\end{flushright}
\vfill
\begin{center}
{\large \bf Chiral Hierarchies, Compositeness \\
and the Renormalization Group}
 \end{center}
  \par \vskip .1in \noindent
\begin{center}
{\bf William A. Bardeen}
  \par \vskip .05in \noindent
{Theoretical Physics, SSC Laboratory \\
 2550  Beckleymeade Ave., Dallas, Texas 75237--3946}
  \par \vskip .05in \noindent
 {\bf Christopher T. Hill and Dirk--U. Jungnickel\footnote{Supported
  by the Deutsche Forschungsgemeinschaft} }
  \par \vskip .05in \noindent
{Fermi National Accelerator Laboratory\\
 P.O. Box 500, Batavia, Illinois, 60510}
  \par \vskip .05in \noindent
\end{center}
\begin{center}{\large Abstract}\end{center}
\par \vskip .05in
\begin{quote}
A wide class of models involve the fine--tuning of
significant hierarchies
between a strong--coupling ``compositeness'' scale, and
a low energy dynamical symmetry breaking scale. We examine the
issue of whether such
hierarchies are generally endangered by Coleman--Weinberg
instabilities. A careful study
using perturbative two--loop renormalization
group methods finds that consistent
large hierarchies are not generally
disallowed.
\end{quote} \vfill
\end{titlepage}

\section{Introduction}

Chivukula, Golden and Simmons \cite{CGS} have recently
examined the question of when it is possible
to tune a large hierarchy in a chiral theory with essentially
composite scalar bosons. This issue can arise in
models such as  heavy--quark condensation models
\cite{Nambu}--\cite{EKR}, models of
broken technicolor \cite{HKOY} or strong
extended technicolor \cite{ATEW}--\cite{MY}.
In ref.~\cite{CGS}, the authors argue that such a hierarchy
between the compositeness
scale $\La$ and the chiral symmetry breaking scale $v$ is
generally endangered by the
Coleman--Weinberg phenomenon \cite{CW}. Quantum fluctuations  drive
the chiral symmetry breaking phase transition to be {\em first} order.
This transition must effectively
be {\em second} order if a large hierarchy
can exist by fine--tuning.  A notable exception to
the general result of ref.~\cite{CGS}
is the single composite electroweak
$I=\hal$ Higgs boson in the top quark condensate
models \cite{BHL}, since there is
only one quartic coupling constant in the minimal version.

This issue goes beyond the statement that large hierarchies
are unnatural because of the fine--tuning of additive quadratically
divergent terms. Even if one can remedy that problem, the authors
of \cite{CGS} argue that such a
fine--tuning is problematic if the
compositeness conditions imply that
some of the coupling constants diverge at
the scale $\La$. If these conclusions
are true then the idea of compositeness
and the existence of a large gauge hierarchy,
$v/\Lambda<<1$, cannot reasonably
go together, in general.

In the present paper we will examine
further this issue raised by \cite{CGS}.
We argue that when compositeness is implemented in a
consistent way, the Coleman--Weinberg
phenomenon does not necessarily arise
and fine--tuning may still be possible.
This conclusion hinges in part upon the use
of the perturbative renormalization group
only in a regime in which it is
valid.  Indeed, in top condensation a la ref.~\cite{BHL},
pains were taken to carefully match the
low energy theory in which the perturbative coupling constant
expansion is valid onto a high energy theory which approaches
the scale $\Lambda$ with a valid nonperturbative dynamics.  We
match a large--$N_c$ expansion near $\Lambda$ onto a perturbative
renormalization group at some scale $\mu_i$.  Typically we choose
$\mu_i/\Lambda \sim 0.05$, but the results are reasonably
insensitive to this choice.

We will reexamine the $U_L(N_f)\times U_R(N_f)$ chiral model of
ref.~\cite{CGS} with elementary fermions having
$N_f$ flavors and $N_c$ gauge degrees of freedom.
Near the composite scale, $\La$, the couplings become large and
nonperturbative methods must be used to analyze the dynamics.
We use the formal large--$N_c$ methods as a guide to determine
the renormalized coupling constants near the composite scale. At
lower scales $\mu\ll\La$, the effective couplings become weak and
the usual perturbative methods become applicable.
The large--$N_c$ couplings are then matched at a scale $\mu_i$
onto the perturbative couplings which satisfy
a two---loop  renormalization group.  Our criterion is that
the two loop terms are no larger than the one--loop terms
at $\mu_i$, and this implies typically $\mu_i\sim 0.05$.
We then evolve to the far infrared.
We search for a Coleman--Weinberg
instability at a scale $v$ which may be intermediate to $\Lambda$
and the far infrared scale of the low energy physics.

We will find using this procedure that
these theories can admit, in general,
large chiral hierarchies. In particular
it turns out that taking into account
two--loop corrections significantly
stabilizes the  $U_L(N_f)\times U_R(N_f)$ models, relative to
the one--loop results. For example, the special case of $N_f=2$,
$N_c$=5, admits a chiral hierarchy
larger than $m_W/m_{\rm pl}$. Furthermore,
in the presence of a strong gauge coupling
like $\alpha_{\rm QCD}$ we do not
find any instability at all.
We will study the stability of these conclusions and
map out a class of models in which
the chiral hierarchies range from $v/\Lambda
\sim 10^{-2}$ to an  arbitrarily infinitesimal $v/\Lambda $. It should be
emphasized that all these results
are physically somewhat  qualitative, and  not
of the form of  rigorous lemmas.  However, once defined precisely, our
procedure leads to rigorous results.

Of course, this does not have a bearing so far as we know on
the origin of gauge hierarchies in nature.  Certainly we would
welcome a {\em raison d'\^etre}\, for these hierarchies, such as,
``if it can exist it must exist''
(see e.g. \cite{HG}). The compositeness
picture at the scale $\Lambda$ suggests
only that hierarchies are associated with the very close proximity
of the high energy coupling parameters to the phase transition boundary.
The approximate recovery of scale invariance in the evolution to
low energies is associated with the tuning of the hierarchy and,
in turn, with the
infrared renormalization group
fixed points \cite{Hill1,HLR} that
accompany the hierarchy.
Perhaps at some deeper level these ingredients alone
can be seen to self--consistently
determine the existence of  hierarchies.

\section{The $U_L(N_f)\times U_R(N_f)$ Higgs--Yukawa model}

In order to illustrate these ideas we take up the model considered in
\cite{CGS} as a generalization of top quark condensate
models. It consists
of $N_f$
left-- and right--handed fermion flavors
$\Psi^j$ ($j=1,\ldots ,N_f$) which
transform in an $N_c$--dimensional representation of some gauge group
and possess a chiral $U_L(N_f)\times U_R(N_f)$ symmetry.
The high--energy dynamics is assumed
to produce a condensate at the scale $\La$ described by the (local)
composite color singlet field $\Si^{ij}\sim {\ol\Psi}_R^j\Psi_L^i$,
whose VEV $v$ may
be interpreted as the order parameter of chiral symmetry breaking. It
transforms in the $(\ol{\bf N}_f ,{\bf N}_f )$ representation of the
chiral symmetry group.
If a large hierarchy between $v$ and the scale $\La$ of new
physics is to be established in a consistent way,
the chiral symmetry breaking phase
transition must be of second order
in the couplings of the high--energy theory
\cite{CCL}. Hence the low--energy
dynamics of the composite Higgs field
$\Si$ and the fermions $\Psi^j$
can be described by an effective Ginsburg--Landau Lagrange
density of the form
\be
 {\cal L} ={\cal L}_{\rm kin} +{\cal L}_{\rm gauge}
 + \frac{\pi y}{\sqrt{N_f}}
 \left( {\ol\Psi}_L \Si\Psi_R +{\rm h.c.}\right)
 -V(\Si ,\Si^\dagger )
 \label{lagrangian}
\ee
where
\be
 V(\Si ,\Si^\dagger )= m^2 \tr\left(\Si^\dagger\Si\right)
 +\frac{\pi^2 \la_1}{3N_f^2}\left(\tr\Si^\dagger \Si\right)^2
 +\frac{\pi^2 \la_2}{3N_f}\tr\left(\Si^\dagger \Si\right)^2
 \label{potential}
\ee
is the most general renormalizable\footnote{For $N_f=4$ there exists an
additional independent renormalizable $U_L(N_f)$ $\times$
$U_R(N_f)$--symmetric term of
the form $\la_3\left(\det\Si +\det\Si^\dagger\right)$. If the dynamics
at $\mu =\La$ is a gauge theory then presumably instantons can lead to
the occurrence of such t'Hooft terms in the low--energy theory.
We will ignore this
possibility for simplicity.}
classical potential symmetric under $U_L(N_f)$ $\times$ $U_R(N_f)$.

As already mentioned, this description of the effective
low--energy dynamics presupposes a chiral symmetry breaking phase
transition of {\em second order} in the coupling constants of the
high--energy theory. However, as has been
pointed out in \cite{CGS}, this can
only be the case in a self--consistent way,
if the phase transition in the
effective low--energy theory (\ref{lagrangian}),
(\ref{potential}) is itself of
second order. This means that its effective
potential in the limit of vanishing
renormalized mass should be minimized
globally at $\Si_c =0$, where $\Si_c$
denotes the {\em classical} scalar field
matrix. Any non--trivial global minimum at $\Si_c\neq 0$ for
$m^2 =0$, i.e. the occurrence of the
Coleman--Weinberg phenomenon \cite{CW},
would clearly indicate a first order
transition of the low--energy theory, therefore rendering the crucial
assumption of a second order transition of the high--energy theory
self--inconsistent. This in turn would
imply that a classically fine--tuned
hierarchy $v/\La$ would always be
destabilized by quantum fluctuations. We
will investigate this potential
problem following the approach of Yamagishi
\cite{Yama}.

The appropriate technical tool to study the
Coleman--Weinberg phenomenon is the
effective potential. Its perturbative
evolution to one--loop for the model
(\ref{lagrangian}), (\ref{potential}) is
given by\footnote{At this stage we
neglect any possible gauge coupling of the
fermions.}
\bea
 V_{\rm eff}&=&\ds{V+\frac{1}{2}\int\frac{d^4p}{(2\pi )^4}
 \tr\ln\left( 1+\frac{W}{p^2}\right)}\nnn
 &-&\ds{\int\frac{d^4p}{(2\pi )^4}\sum_{n=1}^{\iy}
 \frac{(-1)^{n+1}}{n}\left(\frac{\pi y}{\sqrt{N_f} p^2}\right)^n
 \tr\left[ p_\mu\gamma^\mu\left(\Si_c P_R +
 \Si_c^\dagger P_L\right)\right]^n
 +{\rm CT}\; ,}
\eea
where $V$ is the tree--level potential
(\ref{potential}) and $W\equiv W(\la )$
denotes the matrix of second derivatives
of $V$ w.r.t.~the (real) scalar field
components in $\Si$. ${\rm CT}$ is the counter term
for this UV--divergent expression
and $P_{R/L}\equiv (1\pm\gamma_5 )/2$. Using
dimensional regularization and
$\ol{MS}$--renormalization this yields
\bea
 V_{\rm eff} &=& V
 -\ds{\frac{\pi^2}{16}\frac{N_c}{N_f^2}y^4 \left(
 \ln\left(\frac{\pi^2y^2}{N_f}\right)-2u-\frac{3}{2}\right)
 \tr\left(\Si_c^\dagger\Si_c\right)^2 }\nnn
 &-& \ds{\frac{\pi^2}{16}\frac{N_c}{N_f^2}y^4
 \tr\left[\left(\Si_c^\dagger\Si_c\right)^2
 \ln\left(\frac{N_f
 \Si_c^\dagger\Si_c}{\tr\Si_c^\dagger\Si_c}\right)\right]} \nnn
 &+& \ds{\frac{1}{64\pi^2}\tr\left[ W^2\left(\ln\left(\frac{N_f W}
 {\tr\Si_c^\dagger\Si_c}\right)-2u-\frac{3}{2}\right)\right] }
 \label{Veffun}
\eea
 with the scale parameter $u$ defined as
\be
 u=\hal\ln\left(\frac{\tr\Si_{\rm c}^\dagger\Si_{\rm c}}{N_f M^2}\right)\; ,
\ee
and $M$ being the renormalization scale.

The renormalization group equation for this potential may be written as
\be
 {\cal D}V_{\rm eff}(\Si_{\rm c} ,\Si_{\rm c}^\dagger ) = 0
 \label{VRGE}
\ee
with
\be
 {\cal D}=
 -\frac{\prl}{\prl u}+\ol{\beta}_1\frac{\prl}{\prl\la_1}
 +\ol{\beta}_2\frac{\prl}{\prl\la_2}+\ol{\beta}_y\frac{\prl}{\prl y}
 -\ol{\gm}\sum_{i,j}\left(\Si_{\rm c}^{ij}\frac{\prl}{\prl\Si_{\rm c}^{ij}}
 +\Si_{\rm c}^{\dagger ij}\frac{\prl}{\prl\Si_{\rm c}^{\dagger ij}}\right)
 \; ,
\ee
\be
 \ol{\beta}_1 =\frac{\beta_1}{1+\gm}\; ,\;\;
 \ol{\beta}_2 =\frac{\beta_2}{1+\gm}\; ,\;\;
 \ol{\beta}_y =\frac{\beta_y}{1+\gm}\; ,\;\;
 \ol{\gm} =\frac{\gm}{1+\gm}\; ,
 \label{betabar}
\ee
and the $\beta$'s denote the conventional beta functions
\be
 \beta_1 = \mu\frac{d\,\la_1}{d\,\mu}\; ,\;\;\;
 \beta_2 = \mu\frac{d\,\la_2}{d\,\mu}\; ,\;\;\;
 \beta_y = \mu\frac{d\, y}{d\,\mu}\; .
\ee
The anomalous dimension of $\Si$ is defined as $\gm =\hal \mu
(d\ln Z_\Si /d\mu)$, where $Z_\Si$ is the
scalar wave function renormalization constant.

In view of (\ref{Veffun}) the
one--loop or next--to--leading log renormalization
group improved effective potential is then given by
\bea
 V_{\rm eff}^{\rm RG} &=& \ds{\left\{\frac{\pi^2}{3N_f^2}
 \ol{\la}_1 (u,\la )\left(\tr\Si_{\rm c}^\dagger\Si_{\rm
 c}\right)^2
 +\frac{\pi^2}{3N_f}\ol{\la}_2 (u,\la )
 \tr\left(\Si_{\rm c}^\dagger\Si_{\rm c}
 \right)^2\right.  }\nnn
 &-& \ds{\frac{\pi^2}{16}\frac{N_c}{N_f^2}\ol{y}^4 (u,\la )
 \tr\left[\left(\Si_c^\dagger\Si_c\right)^2 \left(
 \ln\left(\frac{\pi^2\ol{y}^2 (u,\la )
 \Si_c^\dagger\Si_c}{\tr\Si_c^\dagger\Si_c}\right)
 -\frac{3}{2}\right)\right] }\nnn
  &+& \ds{\left. \frac{1}{64\pi^2}\tr\left[ W^2(\ol{\la}(u,\la ))
 \left(\ln\left(\frac{N_f W(\ol{\la}(u,\la ))}
 {\tr\Si_c^\dagger\Si_c}\right)-\frac{3}{2}\right)\right] \right\} } \nnn
 &\times& \ds{\exp\left\{ -4\int_{0}^u ds\,
 \ol{\gm}\left(\ol{\la} (s,\la ))\right)\right\}
 }\; .
 \label{Veff}
\eea
The effective coupling constants
$\ol{\la}_1 (u,\la )$, $\ol{\la}_2 (u,\la )$
and $\ol{y} (u,\la )$ are defined by the generically coupled set of
differential equations (the argument $\la$ represents the dependence on
$\la_1$, $\la_2$ and $y$)
\be
 \ol{\beta}_1 (\ol{\la}_1 ,\ol{\la}_2 ,\ol{y} )
 = \frac{\prl\ol{\la}_1}{\prl u}\; ,\;\;\;
 \ol{\beta}_2 (\ol{\la}_1 ,\ol{\la}_2 ,\ol{y} )
 = \frac{\prl\ol{\la}_2}{\prl u}\; ,\;\;\;
 \ol{\beta}_y (\ol{\la}_1 ,\ol{\la}_2 ,\ol{y} )
 = \frac{\prl\ol{y}}{\prl u}
 \label{runningcouplings}
\ee
with boundary conditions $\ol{\la}_1 (0,\la )=\la_1$, $\ol{\la}_2
(0,\la )=\la_2$ and $\ol{y} (0,\la )=y$.

${\rm V}_{\rm eff}$ develops a local minimum in
$\Si_{\rm c}$ different from zero if for some $u_0$ the following two
conditions are fulfilled \cite{Yama}:
\bea \ds{\left.\frac{\prl V_{\rm eff}^{\rm RG}}
 {\prl \Si_c^{ij}}\right|_{u=u_0}}
 &=& \ds{\left.\frac{\prl V_{\rm eff}^{\rm RG}}
 {\prl \Si_c^{\dagger ij}}\right|_{u=u_0}=0}\nnn
 W(u_0) &\equiv& \left.\left(
 \ba{cc}
  \ds{\frac{\prl^2 V_{\rm eff}^{\rm RG}}
  {\prl\Si_c^{ij}\prl\Si_c^{kl}}} &
  \ds{\frac{\prl^2 V_{\rm eff}^{\rm RG}}
  {\prl\Si_c^{ij}\prl\Si_c^{\dagger kl}}} \nnn
  \ds{\frac{\prl^2 V_{\rm eff}^{\rm RG}}
  {\prl\Si_c^{\dagger ij}\prl\Si_c^{kl}}} &
  \ds{\frac{\prl^2 V_{\rm eff}^{\rm RG}}
  {\prl\Si_c^{\dagger ij}\prl\Si_c^{\dagger
  kl}}}
 \ea
 \right)\right|_{u=u_0} >0 \; .
 \label{stabconditions}
\eea
The first relation indicates a stationary
point of ${\rm V}_{\rm eff}^{\rm RG}$,
whereas the second ensures that this stationary
point is indeed a local minimum
of the effective potential. If furthermore
$V_{\rm eff}^{\rm RG}(u_0) < 0$ is satisfied, this
minimum is not only a local but a global one.

Later we will analyze these conditions
in perturbation theory. A consistent
one--loop treatment requires the knowledge of the one--loop
$\ol{\beta}$--functions but only the leading log (or RG--improved
tree--level) terms of (\ref{Veff}).
In this case the conditions
(\ref{stabconditions}) may be rewritten as
\bea
 S(u_0) &\equiv& \ds{\left[4\left(\ol{\la}_1 +\ol{\la}_2\right)
 + \beta_1 (\ol{\la}) +\beta_2 (\ol{\la})\right]_{u=u_0} = 0} \nnn
 P(u_0) &\equiv& \ds{
 \left[4+\beta_1 (\ol{\la})\frac{\prl}{\prl\ol{\la}_1}
 +\beta_2 (\ol{\la})\frac{\prl}{\prl\ol{\la}_2}
 +\beta_y\frac{\prl}{\prl \ol{y}}\right]
 \left(\beta_1 (\ol{\la}) +\beta_2 (\ol{\la})\right)_{u=u_0} >0} \nnn
 \ol{\la}_2 (u_0,\la ) &>& 0 \; .
 \label{stability}
\eea
The condition $V_{\rm eff}^{\rm RG}(u_0) < 0$ yields
\be
 \ol{\la}_1(u_0,\la ) +\ol{\la}_2(u_0,\la ) <0\; .
 \label{absMin}
\ee
Any solution to the condition $S(u_0)=0$ can be understood as a quartic
instability of the effective potential
as is evident in the classical limit, i.e.
for vanishing beta--functions. In addition it allows for a nice
geometrical interpretation in coupling constant space \cite{Yama}:
Since the effective potential (as well as its extrema) is independent of
the renormalization scale $M$ (this is just the meaning of the
RGE (\ref{VRGE})) we may fix it to any physically meaningful value
$M\leq \La$. The flow of
$\ol{\la}_1$, $\ol{\la}_2$ and $\ol{y}$ with $\mu^2\equiv
\tr\left(\Si_c^\dagger\Si_c /N_f\right)$ then determines whether there is an
additional extremum away from $\mu =0$. If for some $u_0 =\ln(\mu_0/M)$ the
flow intersects the ``stability surface'' $S(u)=0$ and in addition the other
conditions in (\ref{stability}) and (\ref{absMin}) are satisfied there is an
absolute minimum of $V_{\rm eff}$ away from zero and the VEV of $\Si$ is
given by
\be
 \frac{1}{\sqrt{2}}
 v_0 \delta^{ij} \equiv\VEV{0|\Si^{ij} |0} =M e^{u_0}\delta^{ij}\; .
\ee

In a second step we will improve the one--loop results by including
two--loop corrections. A consistent treatment then requires the two--loop
corrections to the $\ol{\beta}$--functions
as well as the next--to--leading log
contributions to the RG--improved
effective potential as given in (\ref{Veff}).
Consequently the functions $S(u)$ and
$P(u)$ acquire higher order corrections.

Eqn.~(\ref{absMin}) already shows that the
phase transition of the minimal
top condensation model \cite{BHL},
which has only one quartic coupling
constant, is always of second order.
The compositeness boundary conditions
guarantee that its infrared stable
quasi--fixed point is at $\la >0$ and
(\ref{absMin}) can not be fulfilled.
This is, however, not a general feature
of models with only one quartic coupling
constant, since e.g. scalar QED
clearly exhibits the Coleman--Weinberg
phenomenon \cite{CW}\footnote{It should
be noted that the third condition
(\ref{stability}) is due to the fact that the
model under consideration has two
quartic scalar self--interactions. A similar
condition does not show up in models
with only one quartic coupling like scalar QED.}
and is expected to have a first order phase transition.

\section{An exactly solvable example}

The Lagrange density (\ref{lagrangian}),
(\ref{potential}) may be viewed as the
effective low--energy Landau--Ginsburg description of a gauged
Nambu--Jona-Lasinio model \cite{NJL}.  This is a quantum theory
with a cut--off at the scale $\Lambda$, and should generally be
viewed as an approximation to a more general Lagrangian involving
a series of higher dimension operators:
\be {\cal L}={\cal L}_{\rm kin}+{\cal L}_{\rm gauge}
 + G \left(\ol{\Psi}_L^i \Psi_R^j \right)
     \left(\ol{\Psi}_R^j \Psi_L^i \right)
 \label{NJL-lagr}
\ee
The equivalence is seen by rewriting
(\ref{NJL-lagr}) in a Yukawa form with the help of a static auxiliary
scalar matrix field $\Si$ (see e.g. \cite{Eguchi}):
\be
 {\cal L}={\cal L}_{\rm kin}+{\cal L}_{\rm gauge}
 +y_0 \left(\ol{\Psi}_L \Si\Psi_R +h.c.\right) -m_0^2
 \tr\left(\Si^\dagger \Si\right) \; .
 \label{auxfield}
\ee
The physical low energy effective theory is generated by
integrating out fermion degrees of freedom with momentum
$\mu < p < \Lambda$.  This effective theory cannot contain any physical
implications for physics on scales $p>\La$.
At scales $\mu\ll\La$ the Yukawa interaction
in (\ref{auxfield}) induces the fully gauge invariant,
kinetic and quartic scalar self--interaction terms of (\ref{potential}).
In the fermion bubble approximation, i.e.
to leading order in a large--$N_c$expansion,
$GN_c$ fixed and all gauge couplings
neglected, this model is exactly solvable.
In this approximation the beta functions for the renormalized coupling
constants and the anomalous dimension of the scalar field  are given by:
\be
 \beta_1^{(1/N_c)}=0\; ,\;\;
 \beta_2^{(1/N_c)}=4a\la_2 y^2 -6ay^4\; ,\;\;
 \beta_{y^2}^{(1/N_c)}=2ay^4
\ee
\be
 \gamma^{(1/N_c)}=ay^2
\ee
with $a\equiv N_c/16N_f$. The corresponding
renormalized couplings are
seen to be \cite{BHL}:
\be \la_1 (\mu ) =0\; ,\;\;
    \la_2 (\mu )=\frac{24N_f}{N_c \ln(\frac{\La}{\mu})}
    \; ,\;\;
    y^2 (\mu ) =\frac{8N_f}{N_c \ln(\frac{\La}{\mu})}
    \; ,
    \label{1/nc}
\ee
i.e. $\la_2$ and $y^2$ tend to diverge at
the compositeness scale. We will
always assume in the following that $N_c$ $\geq$ $N_f$. Otherwise the
large--$N_c$ expansion would no longer
be reliable and instead would have to
be replaced by a large--$N_f$
expansion yielding presumably very different infrared dynamics.

For a further discussion of the nature of the chiral symmetry breaking
we have to determine the flow of the coupling constants
$\ol{\la}_1$, $\ol{\la}_2$ and $\ol{y}^2$ which is governed by the beta
functions defined in (\ref{betabar}). The solution to
(\ref{runningcouplings}) can only be given in a transcendental form:
\be \ol{\la}_1 (u)=0\; ,\;\; \ol{\la}_2 (u) =3\ol{y}^2 (u)\; ,\;\;
 2u=\ln\left(\frac{\ol{y}^2 (u)}{y^2}\right)-
 \frac{1}{a}\left(\frac{1}{\ol{y}^2 (u)}-\frac{1}{y^2}\right)\; .
 \label{transcen}
\ee

The full quantum corrections to the effective potential in this
approximation are given by the fermionic one--loop contributions in
(\ref{Veffun}):
\bea
 V_{\rm eff} &=& \ds{\frac{\pi^2
 \la_2}{3N_f}\tr\left(\Si_c^\dagger\Si_c\right)^2
 -\frac{\pi^2}{16}\frac{N_c}{N_f^2}y^4 \left(
 \ln\left(\frac{\pi^2y^2}{N_f}\right)-\frac{3}{2}\right)
 \tr\left(\Si_c^\dagger\Si_c\right)^2 } \nnn
 &-& \ds{\frac{\pi^2}{16}\frac{N_c}{N_f^2}y^4
 \tr\left[\left(\Si_c^\dagger\Si_c\right)^2
 \ln\left(\frac{N_f
 \Si_c^\dagger\Si_c}{M^2}\right)\right]}\; .
 \label{Veff1/N}
\eea
Since this potential is exact to
leading order in $N_c$ it can not be
improved by the RG (to this order).
Hence it should be possible to rewrite
it in a manifestly RG--invariant way.
Indeed, absorbing its $u$--dependence into
the effective coupling constants using
(\ref{transcen}) yields
the desired form of the effective potential:
\bea
 V_{\rm eff} &=& \ds{\left\{\frac{\pi^2
 \ol{\la}_2(u)}{3N_f}\tr\left(\Si_c^\dagger\Si_c\right)^2
 -\frac{\pi^2}{16}\frac{N_c}{N_f^2}\ol{y}^4 (u)\left(
 \ln\left(\frac{\pi^2\ol{y}^2(u)}{N_f}\right)-\frac{3}{2}\right)
 \tr\left(\Si_c^\dagger\Si_c\right)^2\right. } \nnn
 &-& \ds{\left. \frac{\pi^2}{16}\frac{N_c}{N_f^2}\ol{y}^4 (u)
 \tr\left[\left(\Si_c^\dagger\Si_c\right)^2
 \ln\left(\frac{N_f
 \Si_c^\dagger\Si_c}{\tr\Si_c^\dagger\Si_c}\right)\right]\right\}}\nnn
 &\times&\ds{\exp\left\{ -4\int_0^u ds\, \ol{\gamma}(s)\right\} }
\eea
which exactly agrees with the fermionic contribution to (\ref{Veff}).

The stability of this potential may
be investigated be recasting (\ref{Veff1/N})
with the help of (\ref{1/nc}) to become
\be
 V_{\rm eff}=-\frac{\pi^2}{N_f}ay^4
 \tr\left[\left(\Si_c^\dagger\Si_c\right)^2
 \ln\left(\frac{\pi^2y^2\Si_c^\dagger\Si_c}
 {N_f\La^2 e^{3/2}}\right)\right] \; .
 \label{VefflargeNc}
\ee
One sees that the large--$N_c$ fermionic
contribution to the effective potential is stable for
allowed values of the classical field,
$(\tr\Si_c^\dagger\Si )<N_f \La^2$.
For larger values of the classical field,
the expression for the effective potential in
(\ref{VefflargeNc}) is not valid, the fermions actually decouple
and we can not integrate down to the infrared scale.
The apparent instability of (\ref{VefflargeNc})
for large values of the classical field
is completely unphysical.
The critical issue and result is
that we see no instabilities corresponding to
intermediate scales $\mu < \VEV{\Sigma} < \La$.
Thus, the only relevant physical
local minimum remains at the origin, and the theory
is consistently stable against the intermediate Coleman--Weinberg
instability in the large--$N_c$ limit.

\section{The perturbative regime}

Though the fermion bubble approximation
provides a nice method to solve the
model described by (\ref{NJL-lagr}) in
the strong coupling regime, i.e.
close to the scale $\La$, it is a crude
approximation for small $N_c$,
which might give at most some
qualitative physical hints, for the
nonperturbative regime $\mu >\mu_i$
and especially the values of the
coupling constants at $\mu =\mu_i$
once the full theory is considered. Below
$\mu_i$ the  perturbative expansion of the beta
functions
provides a much more accurate
description of how the couplings run with scale. However,
we emphasize that reliable physical
information may not necessarily be obtained
by using only the lowest order
terms of the perturbative beta functions in the renormalization
group equations as done in \cite{CGS}.
At scales near $\La$ the couplings are becoming large
and higher order terms may be essential to determine the evolution.
We will
demonstrate this below numerically by
comparing the one--loop running of the couplings to the two--loop
running at high scales.
At higher scales, the perturbative renormalization
group fails and the couplings must
be matched to the nonperturbative dynamics
near the composite scale such as
provided by the large--$N_c$ running of
$\ol{\la}_1$, $\ol{\la}_2$ and $\ol{y}^2$ to that governed by the
perturbative beta functions at the scale $\mu_i$. Our next task will
therefore be the determination of $\mu_i$ or equivalently the
perturbative regime of coupling constants.

A simple way to obtain a first idea about the regime of couplings
where perturbation theory can be trusted is to derive tree--level
unitarity bounds in the approximation
of vanishing Yukawa coupling $y^2$. The
strongest restrictions are found by considering $\Si^\dagger$ $\Si$
scattering. One obtains the unitarity bounds
\be\ba{rcl}
 \ds{\left|\hal\left( 1+\frac{1}{N_f^2}\right)\la_1 +\la_2\right|}
 &\leq& \ds{\frac{6}{\pi}}\nnn
 \ds{\left|\frac{1}{N_f^2}\la_1\right|} &\leq& \ds{\frac{12}{\pi}}\nnn
 \ds{\left|\frac{1}{N_f^2}\la_1 +\la_2\right|}
 &\leq& \ds{\frac{12}{\pi}}
 \label{unitarity}
\ea\ee
for the singlet--singlet, adjoint--adjoint and singlet--adjoint
channels, re\-spec\-tive\-ly\footnote{
Note that $({\bf N}_f ,\ol{\bf N}_f)\otimes
(\ol{\bf N}_f ,{\bf N}_f)$ $=$ $({\bf 1},{\bf 1})$ $\oplus$
$({\bf 1},{\bf N}_f^2 -{\bf 1})$ $\oplus$ $({\bf N}_f^2 -{\bf 1},
{\bf 1})$ $\oplus$ $({\bf N}_f^2
-{\bf 1},{\bf N}_f^2 -{\bf 1})$ .}. For the
large--$N_c$ boundary condition
$\la_1=0$ this yields in particular $\la_2\leq
1.9$.

Another possibility to determine the perturbative regime in coupling
constant space is to compare the magnitude of the full
$n+1$--loop contribution
to some perturbative quantity to its full $n$--loop contribution. For our
purpose the most natural choice are the beta functions defined in
(\ref{betabar}). The
coupling constants are
certainly outside the perturbative regime, if the
two--loop corrections to their beta functions are bigger than the
one--loop results. Furthermore, since we are interested in using the
perturbative evolution of the
coupling constants as far as possible, i.e.
limiting the crude large--$N_c$
approximation to the smallest possible range
of the scale parameter,
the determination of the full two--loop corrections
to the beta functions is
mandatory in order to minimize perturbative errors.

The one--loop contributions are given by
\be\begin{array}{rclcl}
 \bt_1^{(1)} &=& \ds{\mu\frac{d\, \la_1}{d\, \mu}} &=&
 \ds{\frac{1}{3}\la_1^2 \left(\frac{1}{4}+\frac{1}{N_f^2}\right)
 +\frac{1}{3}\la_1\la_2 +\frac{1}{4}\la_2^2
 +\frac{1}{4}\frac{N_c}{N_f}\la_1 y^2} \nnn
 \bt_2^{(1)} &=& \ds{\mu\frac{d\, \la_2}{d\, \mu}} &=&
 \ds{\frac{1}{6}\la_2^2 +\frac{1}{2N_f^2}\la_1\la_2
 +\frac{1}{4}\frac{N_c}{N_f}\la_2 y^2
 -\frac{3}{8}\frac{N_c}{N_f}y^4} \nnn
 \bt_{y^2}^{(1)} &=& \ds{\mu\frac{d\, y^2}{d\, \mu}} &=&
 \ds{\frac{1}{8}\left( 1+\frac{N_c}{N_f}\right) y^4}\nnn
 \gm^{(1)} &=& \ds{\frac{1}{16}\frac{N_c}{N_f}y^2}\; .
 \label{one--loop}
\eea
Since $\gm^{(1)}\sim O(y^2)$ we have $\ol{\beta}_j^{(1)}$ $=$
$\beta_j^{(1)}$ ($j=1,2,y^2$).

The two--loop corrections may be evaluated using the results of
\cite{MV}. One obtains
\bea
 \bt_1^{(2)} &=&
 \ds{-\frac{1}{24}\left[
 \la_2^3 +\frac{1}{12}\left( 5+\frac{41}{N_f^2}\right)\la_2^2\la_1
 +\frac{11}{3 N_f^2}\la_2\la_1^2
 +\frac{1}{4N_f^2}\left( 3+\frac{7}{N_f^2}\right)\la_1^3 \right] }\nnn
 &-&  \ds{\frac{1}{24}\frac{N_c}{N_f}\left[
 \frac{3}{4}\la_2^2 +\la_2\la_1
 +\left(\frac{1}{4}+\frac{1}{N_f^2}\right)\la_1^2\right] y^2} \nnn
 &+& \ds{\frac{1}{32}\frac{N_c}{N_f}\left[
 \left(\la_2 -\frac{3}{4}\la_1\right) y^4 +\frac{3}{2}y^6\right] }\nnn
 \bt_2^{(2)} &=&
 \ds{-\frac{1}{24}\left[
 \frac{1}{12N_f^2}\left( 5+\frac{41}{N_f^2}\right)\la_1^2\la_2
 +\frac{11}{3N_f^2}\la_1\la_2^2
 +\frac{1}{4}\left( 1+\frac{5}{N_f^2}\right)\la_2^3\right] }\nnn
 &-& \ds{\frac{1}{16}\frac{N_c}{N_f}\left[
 \left(\frac{1}{N_f^2}\la_1\la_2 +\frac{1}{3}\la_2^2\right) y^2
 +\left(\frac{3}{8}\la_2 -\frac{1}{2N_f^2}\la_1\right) y^4
 -\frac{3}{4}y^6\right]}\nnn
 \bt_{y^2}^{(2)} &=&
 \ds{\frac{1}{576}\left[\frac{1}{N_f^2}
 \left( 1+\frac{1}{N_f^2}\right)\la_1^2
 +\frac{4}{N_f^2}\la_1\la_2
 +\left( 1+\frac{1}{N_f^2}\right)\la_2^2\right] y^2} \nnn
 &-& \ds{\frac{1}{48}\left[\frac{1}{N_f^2}\la_1
 +\hal\left( 1+\frac{1}{N_f^2}\right)\la_2\right] y^4
 -\frac{1}{64}\left[
 \frac{1}{8}+\frac{3}{2}\frac{N_c}{N_f}-\frac{1}{N_f^2}\right] y^6 }\; .
 \label{two--loop}
\eea
For simplicity we have neglected the effects of gauge couplings in
these expressions. This may only be
justified at high scales where they
may be assumed to be small. We will
comment later on the modifications induced by a sizable gauge
coupling, e.g. $\alpha_{\rm QCD}$, on the running of $\ol{y}^2$,
$\ol{\la}_1$ and $\ol{\la}_2$.

Since $\gm^{(2)}$ is of the same order in the coupling constants as
the two--loop contributions to the beta functions we find at the
two--loop level
\be
 \ol{\beta}_j =\left[ 1-\gm^{(1)}\right]\beta_j^{(1)}
 +\beta_j^{(2)}\; ;\;\;\; j=1,2,y^2\; .
\ee

Hence
the evolution of $\ol{\la}_1$ close to the compositeness boundary
conditions ($\ol{\la}_1=0$, $\ol{y}^2=\ol{\la}_2/3$)
to two loops is dominated by
\be
 \left.\ol{\beta}_1\right|_{\ol{\la}_1=0,\;\;\ol{y}^2=\ol{\la}_2 /3}
 =\frac{1}{4}\ol{\la}_2^2 -\frac{1}{24}
 \left[ 1+\frac{1}{4}\frac{N_c}{N_f}\right]\ol{\la}_2^3\; .
 \label{l1=0}
\ee
The positive one--loop contribution to $\ol{\beta}_1$
drives $\ol{\la}_1$ negative as one evolves
downwards from $\mu_i$.
This is qualitatively changed if the absolute value
of the (negative) two--loop contribution becomes larger than the
one--loop result. $\ol{\la}_1$ is then driven into the positive region
as displayed in Fig.~\ref{CGSplot}
and for even larger initial values of $\ol{\la}_2$
both scalar self--interaction couplings develop an infrared
singularity. Fig.~\ref{CGSplot} furthermore shows that the
low--energy physics of this model is
fairly insensitive to the exact
initial values especially if they are large
and the two--loop evolution of the coupling constants is used. This
fact is due to the existence of an effective infrared fixed point in
coupling constant space \cite{Hill1,HLR}.

In view of (\ref{l1=0}) we therefore conclude that
the perturbative running of the coupling constants with large--$N_c$
initial values is definitely misleading
if $\ol{\la}_2$ is larger than
\be
 \ol{\la}_{2i}^{\rm\; pert}\equiv \frac{24N_f}{4N_f+N_c}
\ee
since then the perturbative
error is about $100\%$. In fact, in order to
avoid large perturbative errors
especially in the high $t$--regime
($t=\ln(\La /\mu )$),
we have to choose a starting value
$\ol{\la}_{2i}\equiv\ol{\la}_2(\mu_i)$
which is somewhat smaller than
$\ol{\la}_{2i}^{\rm\; pert}$. On the
other hand, one should not use the
large--$N_c$ running of the couplings for
too many orders of magnitudes
in the scale parameter. In order to obtain
the smallest possible numerical
errors in our calculation we will therefore
match the perturbative to the
large--$N_c$ running of the coupling
constants at a scale $\mu_i$ for which
\be
 \hal\ol{\la}_{2i}^{\rm\; pert}
 < \ol{\la}_{2i} < \ol{\la}_{2i}^{\rm\; pert}
 \label{matchingpoint}
\ee
or equivalently
\be
 1+4\frac{N_f}{N_c}
 < \ln\left(\frac{\La}{\mu_i}\right) <
 2\left( 1+4\frac{N_f}{N_c}\right)\; .
\ee
This means that
for e.g. $N_f =2$, $N_c =3\; (5)$ one has to
use the $1/N_c$--running for
approximately $1.6$ ($1.1$) orders of magnitude,
in order to reach the matching
scale $\mu_i$. If one prefers however, to
chose a matching scale $\mu_i$
at which the perturbative error is reduced to
about $50\%$ as for the lower
bound of (\ref{matchingpoint}) one has to use
the large--$N_c$ running for $3.2$ ($2.2$) orders of magnitude.

One might feel somewhat
uneasy about using the bubble approximation over such a large
range in the scale parameter
especially for small $N_c$. We have paid regard
to this concern in our numerical analysis by varying the
initial values of the coupling constants considerably around their
large--$N_c$ values at $\mu_i$.
On the other hand
it is also known that higher
dimensional operators in the NJL--Lagrangian, which can be determined
once the precise high--energy dynamics above the scale
$\La$ is known,
can also change the compositeness
conditions \cite{Bardeen}. $\ol{\la}_2$ generically takes on a large but
{\em finite} value at $\La$ once these operators are taken into
account. This
will result in a somewhat shorter and
perhaps even faster evolution down to
$\ol{\la}_{2i}^{\rm\; pert}$. Furthermore we expect subleading
corrections to the large $N_c$--running close to $\La$ to have similar
effects.

\section{Numerical results}

For physical applications it is
important to establish the range of hierarchies
permitted by a given model. Not only is it
essential to determine
if the evolution of the coupling constants
intersects the stability
surface, but it is also to establish
how many orders of magnitude
one can run before this
happens, i.e. how large a hierarchy $v/\La$ one can establish
without running into a Coleman--Weinberg instability. In addition one
would like to know how this depends on $N_f$ and $N_c$.
Finally one should investigate how sensitive the evolution and its
stability are w.r.t. a
modification in the running of
$\ol{y}^2$, e.g. due to the in fluence
of a fairly strong gauge interaction like $SU(3)_{\rm QCD}$.
We will address these issues in the following.

To get a first idea of how the
coupling constants evolve with scale one can
(as done in \cite{CGS}) make the
simplifying assumption of a constant Yukawa
coupling, i.e. $\ol{y}^2=1$. In this case
the stability hyper--surface is reduced to a
line. In Fig.~\ref{CGSplot} we have
plotted the one-- and two--loop perturbative
flow in coupling constant space for this situation for
various large--$N_c$ initial values and $N_f=2$, $N_c=3$ including
this stability line. The perturbative results are trustworthy
up to an initial value of
$\ol{\la}_{2i}^{\rm pert}$ $\approx$ $4.6$ for for this
case. Our numerical results agree
with those of \cite{CGS}, though, as explained
before, our interpretation is different.
The instabilities found in \cite{CGS} all
correspond to initial values of the
coupling constants for which perturbation
theory breaks down. We therefore
choose to apply the large--$N_c$ approximation to
model the dynamics in the strong--coupling
regime until perturbative values are
reached. Taking these as initial values
for the perturbative evolution the
RG--trajectories {\em never} cross the
stability line therefore indicating a phase
transition of {\em second} order.
For larger initial values the two--loop evolution
clearly shows that the perturbative
results can no longer be trusted. We therefore
conclude that compositeness boundary
conditions may well be compatible with a
second order phase transition.
Furthermore one notices from Fig.~\ref{CGSplot}
that the two--loop evolution greatly improves
the stability of the model over the one--loop
running. This may however be a model--dependent result. Another
feature of the numerical analysis is
that the stability increases with increasing
$N_c$ for fixed $N_f$. This is,
however, to be expected from our large--$N_c$
analysis of section III.

A constant $\ol{y}^2$ does of course not correspond to a
physically motivated model.
The simplest assumption corresponding to a real
model is that of a vanishing Yukawa coupling.
In this case our methods
clearly signal a {\em first} order
phase transition independent of the values of
$N_f$ and $N_c$ for a wide range of
perturbative initial values for the coupling
constants. This agrees with the results
of \cite{Paterson} and \cite{Shen} and is
not too surprising because of the
lack of infrared fixed points in coupling
constant space. For our purpose,
however, this situation is only of little
interest, since a vanishing
Yukawa coupling is not consistent with the
compositeness conditions we wish to investigate.

We will therefore turn to the full
perturbative running of $\ol{y}^2$ as given by
(\ref{one--loop}) and (\ref{two--loop}).
The flow of the coupling constants
resembles that of Fig.~\ref{CGSplot}.
In Figs.~\ref{S23} and \ref{S25} we have
plotted the perturbative evolution of the stability
function $S(u)$ (as defined in
(\ref{stability}) for a leading log analysis)
for the cases $N_f=2$, $N_c=3$ and
$N_f=2$, $N_c=5$, respectively, for the leading as well as for
the next--to--leading log expression of
the effective potential and for various
large--$N_c$ motivated initial values of
the couplings inside the perturbative
regime. In the case of $N_c=3$ the
function $S$ generally exhibits a zero which
can be seen to correspond to a local
minimum of $V_{\rm eff}^{\rm RG}$ indicating a
first order phase transition. However,
one should notice that the zeros occur
after a significant amount of running
therefore allowing for the tuning of
hierarchies of approximately
$\La /v\approx 10^{10}$ if the two--loop evolution is
used. Again we observe that the
NLL--corrections as well as an increasing
$N_c$ seem to stabilize the potential
for this particular model considerably.

Though the situation of a constant
$\ol{y}^2$ considered earlier is not fully
realistic, it mimics the low--scale
behavior of the Yukawa coupling in the
presence of a strong asymptotically
free gauge coupling. Such a coupling would
significantly affect the running of
$\ol{y}^2$ at lower scales leading to an
effective infrared fixed point
\cite{Hill1,HLR} which in turn might
suggests, that the phase transition is of second order.
This is of course also a more realistic
configuration for physical applications in
heavy quark condensation or strong ETC models.

For definiteness we will assume in the
following that the gauge group is $SU(N_c)$ and that the fermions
transform in the fundamental representation.
The beta function for $\ol{y}^2$ at the
one--loop level is then given by
\bea
 \bt_{y^2}^{(1)} &=&
 \ds{\frac{1}{8}\left( 1+\frac{N_c}{N_f}\right) y^4
 -\frac{3(N_c^2-1)}{2\pi N_c}\alpha_s y^2}\nnn
 \beta_{\alpha_s}^{(1)} &=& \ds{-\frac{1}{\pi}\left(
 \frac{11}{6}N_c -\frac{1}{3}N_f\right)\alpha_s^2}
 \label{y^2gauge}
\eea
with $\alpha_s =g^2 /4\pi$. The effect of the additional term for
$\beta_{y^2}$ is certainly small at high scales due to asymptotic
freedom (for not too large $N_f$). In the infrared however, it becomes
important and effectively stops the running of $y^2$ due to its
negative sign. This situation is hence somewhere in between the running
of $\ol{y}^2$ according to
(\ref{one--loop}) and a constant Yukawa coupling. We
therefore expect a somewhat more stable evolution of $\ol{\la}_1$ and
$\ol{\la}_2$.

In Figs.~\ref{GS23} and \ref{GS25}
we have plotted the function $S(t)$ for
$N_f=2$, $N_c=3$ and $N_f=2$, $N_c=5$,
respectively, using various large--$N_c$
motivated initial values
$\ol{\la}_{2i}\leq\ol{\la}_{2i}^{\rm pert}$. We have
normalized $\alpha$ to be approximately
of the size of the QCD--coupling, i.e.
$\alpha_s (m_Z)=0.1$. Because of
its smallness, especially at high scales, we have included only
corrections of order $\alpha$ in the beta functions, i.e. the
additional term in (\ref{y^2gauge}). For further definiteness we have
chosen $\La$ $=$ $m_{\rm Pl}$ $\approx$ $10^{19}{\rm GeV}$.

Figs.~\ref{GS23} and \ref{GS25}
show that the two--loop evolution of the couplings
never crosses the stability surface $S(t)=0$. Hence
one can self--consistently fine--tune a hierarchy
between the weak and even the Planck scale without running into a
Coleman--Weinberg instability. Remarkably, at comparably high initial
values for $\ol{\la}_{2i}$ for which the one--loop evolution signals a
potential instability, the situation is improved considerably by the
NLL corrections. We have checked that these statements are
insensitive to changes in $N_f$ and $N_c$
as long as $N_f$ is not much larger
than $N_c$. In particular we find, as expected
by our large--$N_c$ analysis and
demonstrated in Fig.~\ref{GS25}, that the
stability increases again with
increasing $N_c$ for $N_f$ held fixed.

Finally we should mention that we
have checked the sensitivity of our numerical
analysis w.r.t. small deviations from the large--$N_c$ initial
values for $\ol{\la}_1$ and $\ol{y}^2$
according to a parameterization
\be
 \ol{\la}_{1i}=a\; ,\;\;\;
 \ol{y}^2_i =\frac{1}{3}\ol{\la}_{2i}\left[ 1+b\right]\; .
\ee
We find that for reasonably small values
of $a$ and $b$, i.e. $|a|, |b|\leq 0.3$,
our quantitative results are fairly
insensitive therefore not altering our
qualitative statements.

The establishment of large
hierarchies by fine--tuning can be very model
dependent as emphasized by the authors
of \cite{CGS}. Specific models may
require careful analysis to establish
the range of hierarchies which may be
achieved by the dynamics. Even models
with several effective couplings at low
energy may be able to support large
hierarchies as in the examples studied in
this paper.

\section{Conclusions}

In this paper we have argued that
fine--tuned chiral hierarchies and
compositeness boundary conditions on coupling
constants can go together in a
self--consistent way, avoiding the
general problem of intermediate Coleman--Weinberg
instabilities. Our analysis has focused on a chiral
$U_L(N_f)\times U_R(N_f)$--symmetric
Yukawa--Higgs--model, though the
methods applied are more general.
By calculating the renormalization group improved
effective potential including
all next--to--leading log contributions,
and implementing the matching of the perturbative
running to the large--$N_c$ evolution, we are able
to provide numerical evidence that fine--tuned
hierarchies are not endangered
by the Coleman--Weinberg instability. We
carefully
match the nonperturbative,
large--$N_c$ running of the coupling
constants, as they become large when approaching the
compositeness scale $\La$,
to their two--loop perturbative RG--evolution
within a valid range of applicability. This allows us to
obtain reliable results for
the full range of momentum scales, up to the
compositeness scale. Curiously,
the phenomenologically most relevant case,
that of a strong non--abelian gauge
coupling to the fermions, allows fine--tuning
hierarchies as large as
$m_W/m_{\rm pl}$, owing to the presence of an effective
nontrivial infrared fixed point.
We find generally that the
next--to--leading log corrections improve
the stability over the leading
log or one--loop results. Furthermore,  the
stability of the hierarchy generally
increases significantly with growing $N_c$.

\vspace{.5 cm}
\ \\
\bf{Acknowledgment}: \normalsize D.--U.~J. is grateful to the
SSC--Laboratory for its hospitality during the final stage of this work.
We thank S. Chivukula for stimulating discussions.
\vspace{ 1 cm}

\newpage

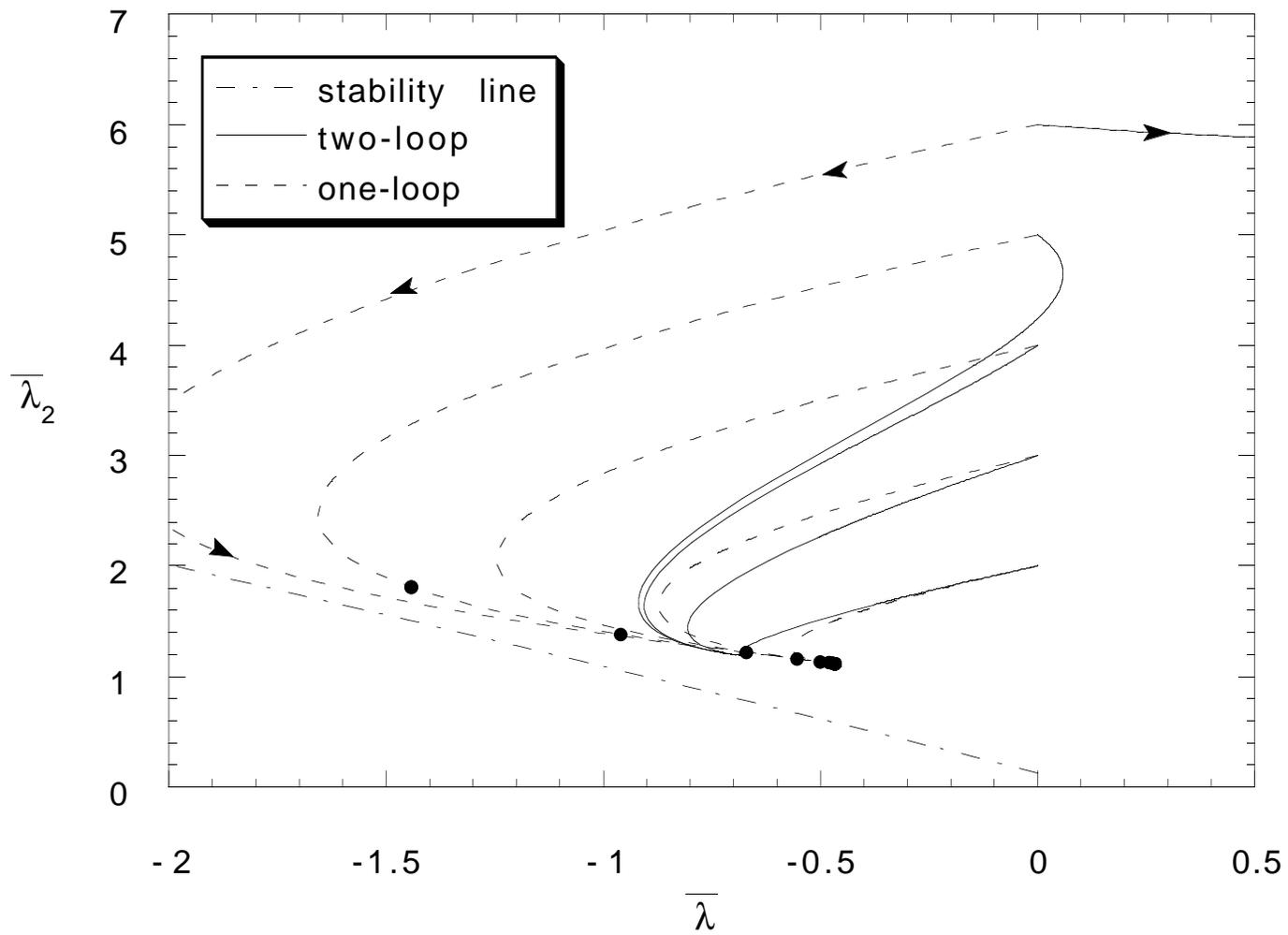
\begin{figure}
 \caption{\em Perturbative one-- and two--loop renormalization group
 trajectories in
 $\ol{\la}_1$--$\ol{\la}_2$ space for
 $N_f=2$, $N_c=3$ and $\ol{y}^2=1$. The evolution is
 displayed for a running of 18 orders of magnitude in
 $\mu$ for several large--$N_c$ initial values. The arrows point in the
 direction of the flow with decreasing scale. Each dot in one of the
 curves indicates an evolution by one order of magnitude. }
 \label{CGSplot}
\end{figure}

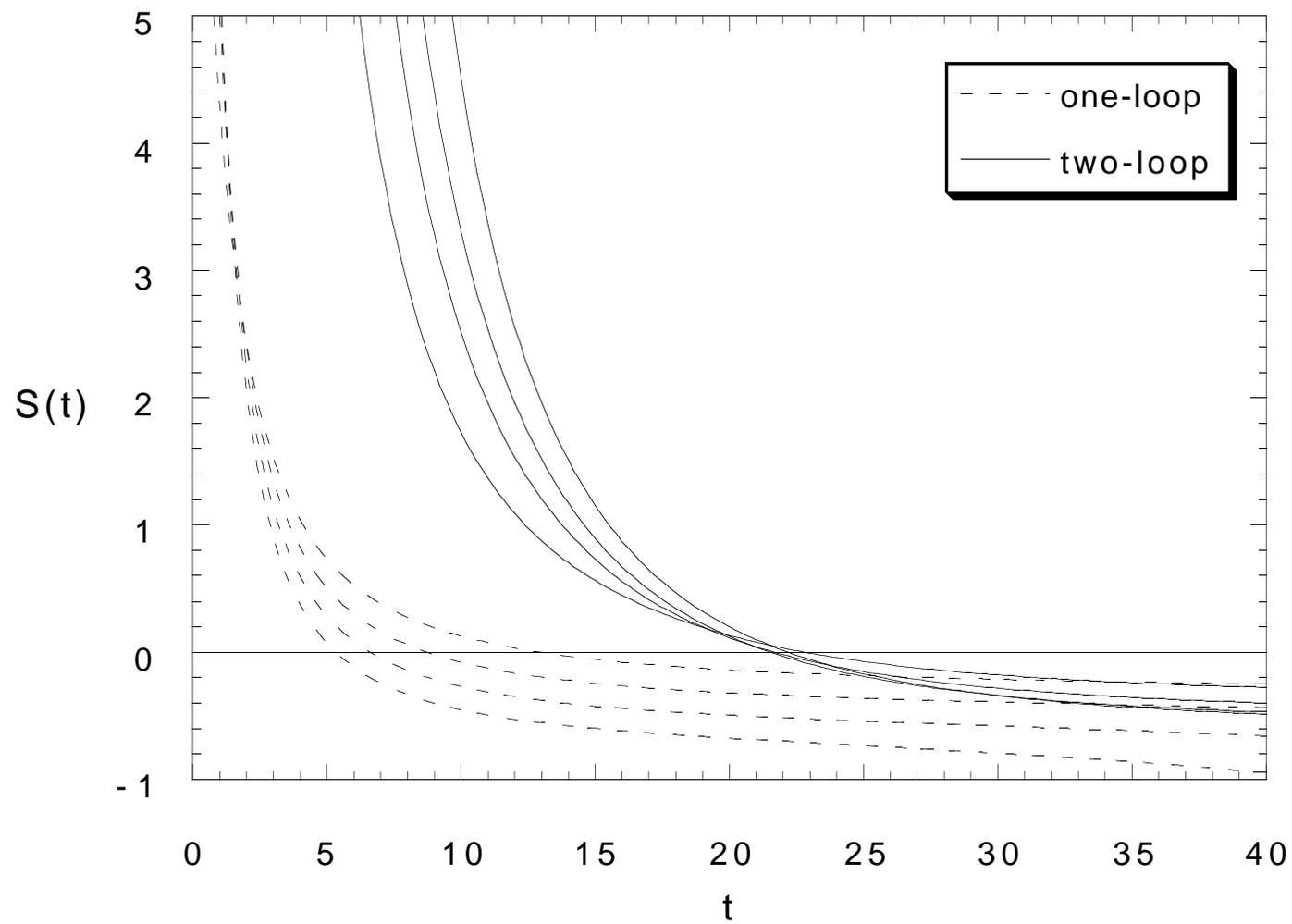
\begin{figure}
 \caption{\em This Figure shows the one-- and two--loop evolution of the
 stability
 function $S(t)$ with $t$ without
 gauge coupling for $N_f=2$, $N_c=3$ and the
 full running of $\ol{y}^2$.}
 \label{S23}
\end{figure}

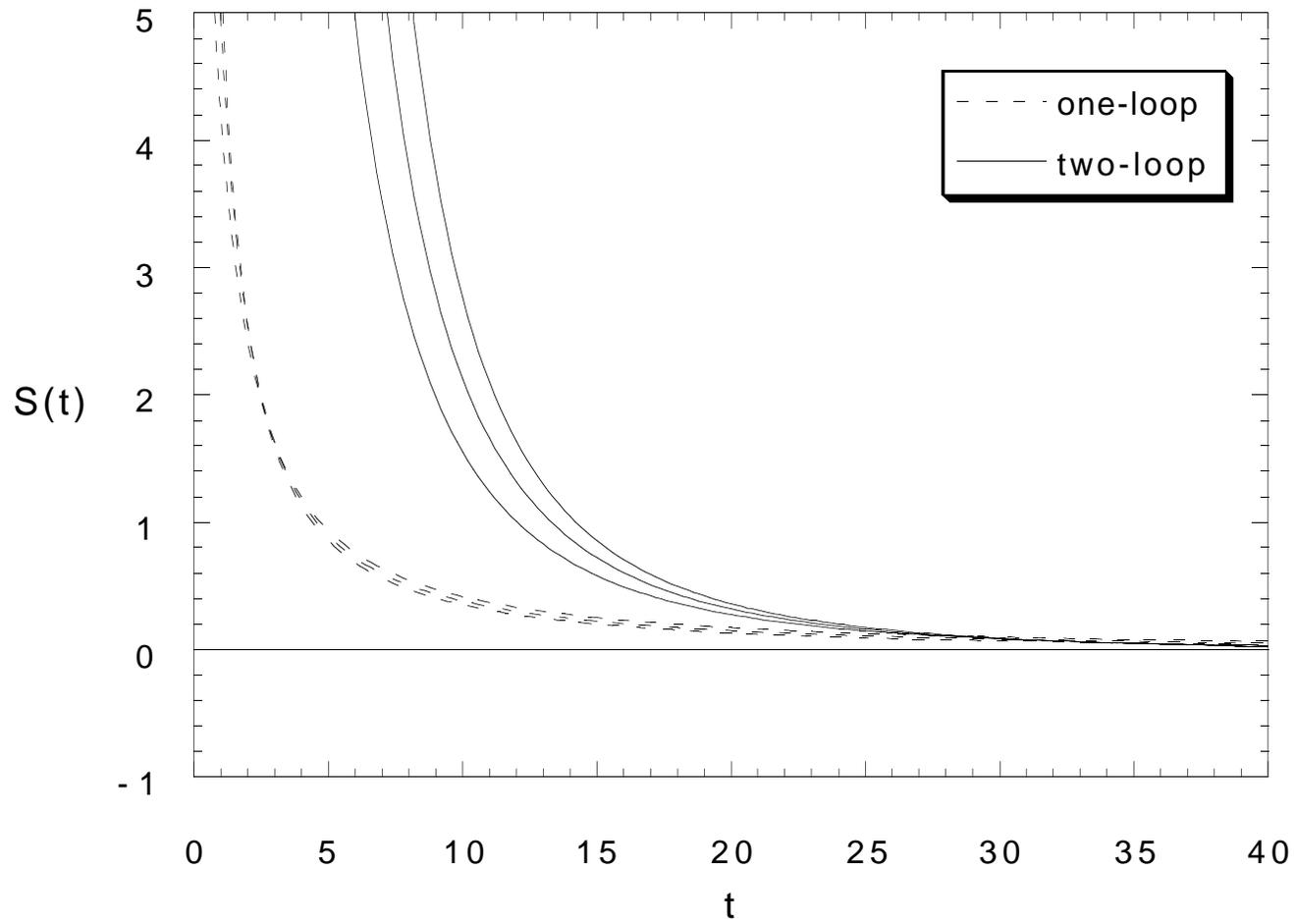
\begin{figure}
 \caption{\em The same as Figure 2 but with increased number of fermion
 colors $N_c=5$.}
 \label{S25}
\end{figure}

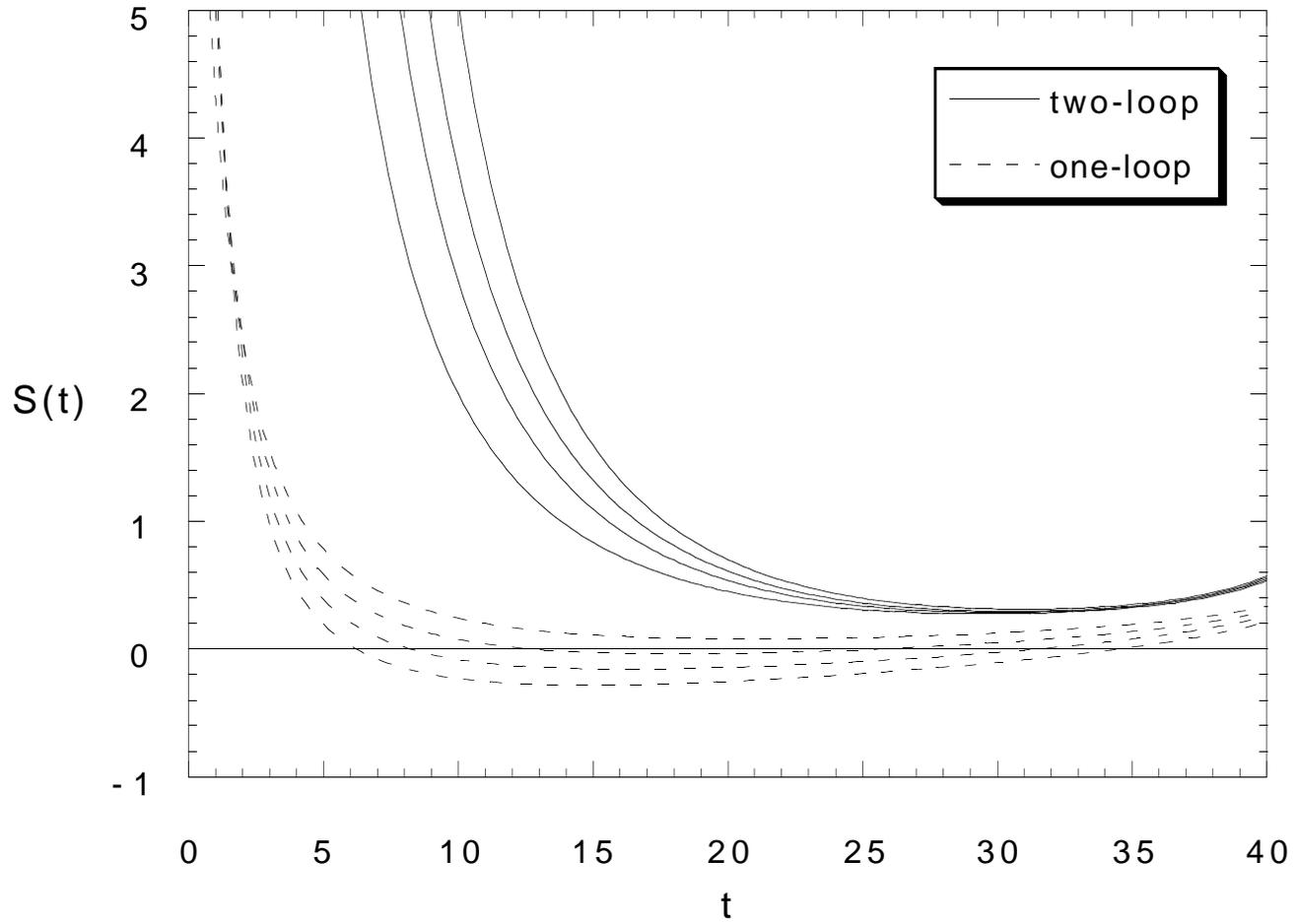
\begin{figure}
 \caption{\em One-- and two--loop evolution of the
 stability
 function $S(t)$ with $t$ for $N_f=2$ and
 $N_c=3$ and the running of $\ol{y}^2$
 modified by a strong gauge coupling $\alpha_s$.}
 \label{GS23}
\end{figure}

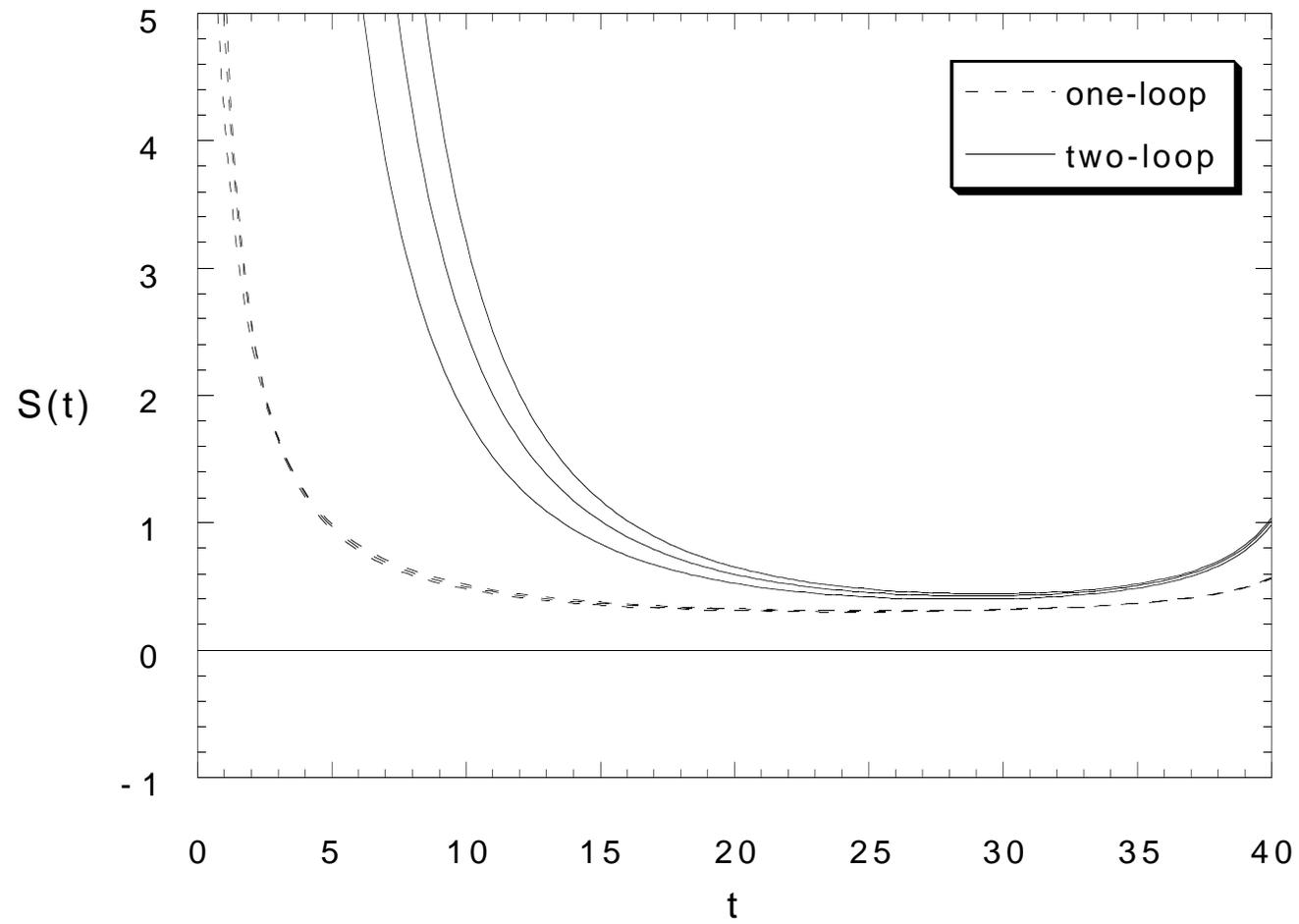
\begin{figure}
 \caption{\em The same as Figure 3 but with increased number of fermion
 colors $N_c=5$.}
 \label{GS25}
\end{figure}

\end{document}